# Magnetic and transport anomalies and large magnetocaloric effect in cubic $R_4$PtAl ($R$ = Ho and Er)

Kartik K. Iyer [1, 2,*], Sudhindra Rayaprol [3], Ram Kumar [1, 4] Shidaling Matteppanavar [2], Suneel Dodamani[2], Kalobaran Maiti [1] and Echur V. Sampathkumaran [6, *]

[1]Tata Institute of Fundamental Research, Homi Bhabha Road, Colaba, Mumbai – 400005, India
[2]KLE Society's Dr. Prabhakar Kore Basic Science Research Centre, KLE Academy of Higher Education and Research, Belagavi- 590010, India
[3]UGC-DAE-Consortium for Scientific Research -Mumbai Centre, BARC Campus, Trombay, Mumbai – 400085, India
[4]Maryland Quantum Materials Center and Department of Physics, University of Maryland, College Park, Maryland 20742, USA
[5]KLE Society's Basavaprabhu Kore Arts, Science & Commerce College, Chikodi 591201, India
[6]Homi Bhabha Centre for Science Education, TIFR, V. N. Purav Marg, Mankhurd, Mumbai – 400088, India

**\*** Correspondence: iyer@tifr.res.in; sampathev@gmail.com

## Abstract

We report the electronic properties of $R_4$PtAl ($R$ = Ho, and Er), which contains 3 sites for R, by the measurements of magnetization (ac and dc), heat-capacity, transport, and magnetoresistance (MR). Dc magnetization data reveal antiferromagnetic order below 19 K and 12 K in Ho and Er compounds, respectively. Additional features observed at lower temperatures (12 K for Ho$_4$PtAl and 5 K for Er$_4$PtAl) are akin to cluster spin-glass phase. Resistivity data exhibit a weak minimum at a temperature marginally higher than their respective Néel temperature ($T_N$) which is unusual for such rare-earths with well-localized 4$f$ states. Isothermal magnetization and magnetoresistance data well below $T_N$ exhibit signatures of a subtle field-induced magnetic transition for a small magnetic field (<10 kOe). Notably, the isothermal entropy change at $T_N$ has the largest peak value within this rare-earth family; for a field change from zero to 50 kOe, the entropy change is ~14.5 J/kg K (Ho4PtAl) and ~21.5 J/kg K (Er$_4$PtAl) suggesting a role of anisotropy of 4$f$ orbital in determining this large value. The results provide some clues for the advancement of the field of magnetocaloric effect. The magnetocaloric property of Er$_4$PtAl is nonhysteretic meeting a challenge to find materials with reversible magnetocaloric effect.

## 1. Introduction

There is a general consensus that the technology of 'magnetic refrigeration' can increase the efficiency of cooling when compared with conventional gas-compression route. Therefore, there are constant efforts in the current literature to find clues to improve the magnetocaloric effect (MCE) as well as to discover new solid materials which are cost-efficient as well as those minimize the problems due to irreversibility in magnetic-field ($H$) cycling. Gd metal, being a large magnetic-moment element, seems to remain as the benchmark for applications, even after several decades of intense research. Needless to emphasize that the exceptional MCE behavior of $Gd_5Si_2Ge_2$ reported by Pecharsky and Gschneidner [1] in 1997 has been at the centerstage of the field of MCE, but its application has been limited due to hysteretic nature of the first-order magnetic-field induced magnetic transition. Naturally, the search for new rare-earth ($R$) compounds with magnetic anomalies continues to be an important direction of research to find favorable MCE materials in different temperature ($T$) ranges. Keeping this scenario in mind, we have investigated the magnetic and MCE behavior of two rare-earth compounds, $R_4PtAl$ ($R$ = Ho and Er).

The compounds of the type $R_4TX$ ($T$ = transition metals and $X$ = p-block metal) with the cubic $Gd_4RhIn$-type structure [2-5] provided a new platform in the field of 'frustrated magnetism', as multiple sites for $R$ seem to favor spin-glass state, competing with antiferromagnetism (AF) and ferromagnetism (F) [6-11]. The crystal structure is shown in Fig. 1 to bring out the surrounding of the three $R$ ions. Complexities of crystallography for this family have been discussed by Engelbert and Janka in the literature [5] and also presented in our earlier publication [7].

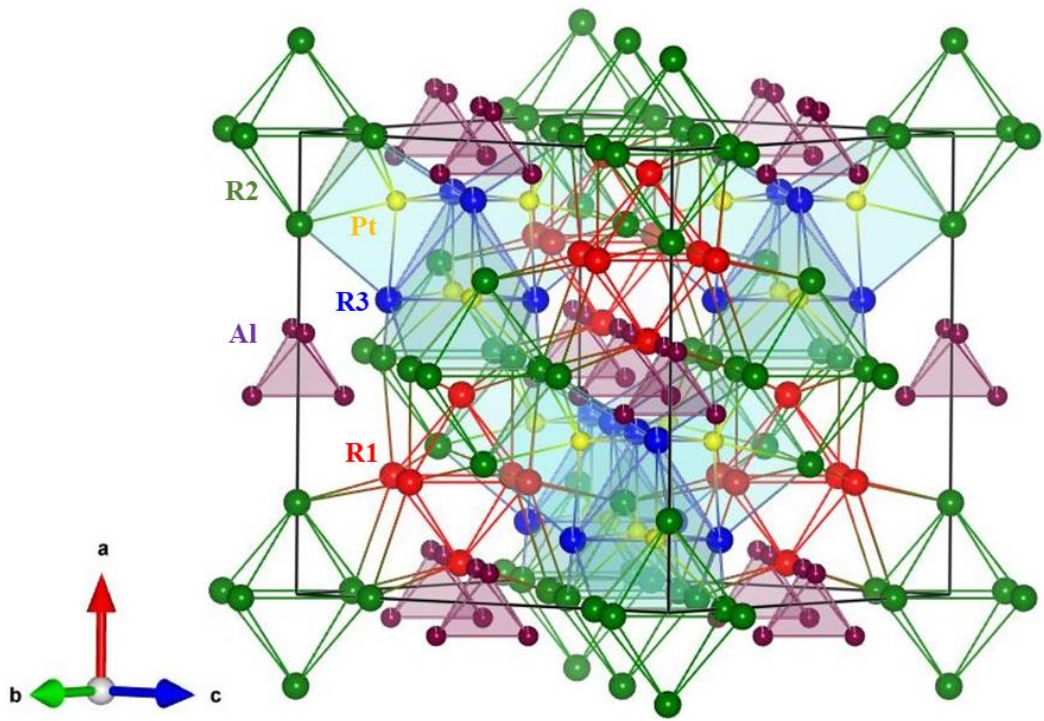

**Figure 1**. Crystal structure of $R_4PtAl$ ($R$= Rare-earth). The rare-earths with different chemical environments can be seen. The rare-earths $R$1 (red) and $R$2 (green) form separate octahedra and $R$2 and $R$3 form an octahedra with Pt (yellow) at the centre. The rare-earth $R$3 (blue) forms a tetrahedra. Al forms a tetrahedra.

Recent investigations on some members of this class of compounds have established that there are novel transport anomalies, not only in the magnetically ordered state, but also in the paramagnetic state [6-11]. Here, we focus on the heavy $R$ members of the Pt-based family, $R_4$PtAl, where the 4$f$ states are far away from the Fermi level (thereby avoiding complex well-known phenomena due to 4$f$-hybridization in light rare-earths); Pt 5$d$ electrons are expected to be weakly correlated, however possessing strong spin-orbit coupling. Extensive studies in recent years revealed that Gd$_4$PtAl shows re-entrant spin-glass behavior around 20 K below its Néel temperature ($T_N$ = 64 K) [6], whereas Tb$_4$PtAl exhibits spin-glass features at the onset of the AF order at 50 K and an additional spin-glass anomaly at a further lower temperature [7]. Curiously, Dy analogue undergoes a ferromagnetic transition at ($T_C$ =) 32.6 K indicating an unusual role of Pt 5$d$ states on the magnetism of a rare-earth with well-localized 4$f$ orbital [11], which transforms to spin-glass phase around 20 K [8]. There are magnetic-field induced features as well, attributable to first-order transitions in most of these compounds. The experimental results presented in this article on Ho and Er compounds reveal fascinating magnetic and transport anomalies with an exceptional MCE, as inferred from the values of the isothermal entropy change (ΔS), in comparison with the values reported earlier for this family; this provides some clues for further theoretical advancement in the field of MCE to enable the discovery of new materials for room temperature applications.

## 2. Materials and methods

Samples were prepared in polycrystalline form by arc melting stoichiometric amounts of constituent elements in argon atmosphere. The Er compound needed to be annealed at 650º C for 8 days. Powder x-ray diffraction patterns (XRD) shown in Fig. 2 were obtained using Cu-Kα radiation and these were found to be in good agreement with those reported in Ref. 5. Rietveld refinement further helped in understanding the diffraction patterns. Such an analysis enabled us to ascertain the formation of the cubic phase with the space group $F\bar{4}3m$. It may be mentioned that XRD pattern showed a weak extra line around 36º for the unannealed Er sample, which disappeared after annealing. Scanning electron microscopic images revealed homogeneity of the samples without showing any extra phase. Dc magnetic susceptibility ($\chi$) (1.8 to 300 K) and isothermal magnetization ($M$) measurements were performed with the help of a commercial (Quantum Design) superconducting quantum interference device; ac $\chi$ data were also collected in a $T$-region of interest with an ac field of 1 Oe with different frequencies ($\upsilon$ = 1.3, 13 and 133 Hz) with the same magnetometer. Heat-capacity ($C$) and electrical resistivity ($\rho$) and magnetoresistance (MR) measurements down to 1.8 K and also as a function of $H$ were caried out using a commercial (Quantum Design Physical Property Measurement System). Unless otherwise stated, all the measurements were performed for the zero-field-cooled (ZFC) condition of the specimens.

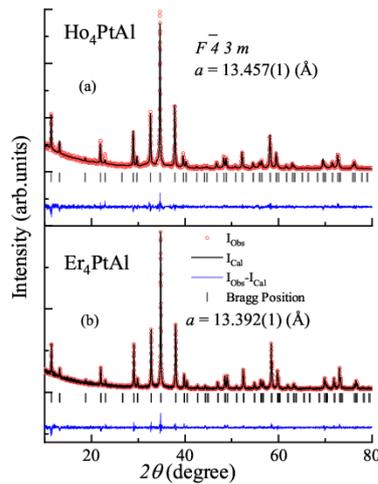

**Figure 2**. X-ray diffraction patterns of Ho$_4$PtAl and Er$_4$PtAl at room temperature, along with Rietveld fitting results.

## 3. Results and discussions

Figs. 3a-b and 4a-b show χ(T) measured in 5 kOe magnetic field as well as in 100 Oe field for both the samples. Inverse χ, plotted for 5 kOe data in the mainframe, is linear above 50 K.

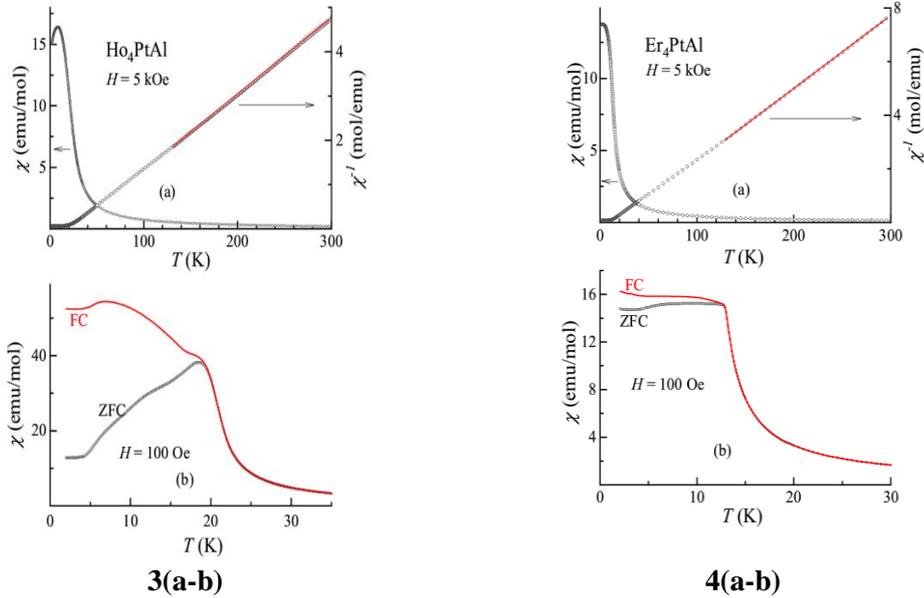

**3(a-b)**      **4(a-b)**

**Figure 3 and Figure 4.** The temperature dependent dc magnetic susceptibility and inverse susceptibility for $Ho_4PtAl$ and $Er_4PtAl$, respectively, measured in (a) 5 kOe and (b) 100 Oe. In (b), the curve obtained for field-cooled warming condition is also included. In (a), a straight line in the inverse χ plot above 120 K represents Curie-Weiss fitting.

The values of the effective magnetic moments (~10.7 $\mu_B$ and 9.82 $\mu_B$ per Ho and Er, respectively) derived from the high-temperature paramagnetic region are in good agreement with the theoretical values. The corresponding values of paramagnetic Curie temperature ($\theta_p$) are ~27 K and 16 K. The positive sign of $\theta_p$ indicates that ferromagnetic coupling between the rare-earth moments is dominant in these compounds in the Curie-Weiss regime, as also observed in the cases of Gd, Tb and Dy members [6-8]; the situation is more complex at lower temperatures as described below. For the Ho compound, as the $T$ is lowered, there is a peak in χ(T) in the ZFC curve for $H = 100$ Oe near 19 K (Fig. 3b) attributable to long range antiferromagnetic order, followed by a shoulder near 12 K and a flattening below about 5 K, as though there are additional magnetic transitions. The ZFC curve obtained in 5 kOe field smears out these additional magnetic features. This suggests that the magnetism is sensitive to external magnetic fields. Irreversibility appears in the 100 Oe field-cooled (FC) and ZFC curves (Fig. 3b) around 19 K, signaling a possibility of spin-glass phase, though other factors (like domain wall boundary effects) have also been known to result in such a feature in ferromagnetic and antiferromagnetic materials. In the case of Er sample, the features appear similar with magnetic ordering (presumably of an AF type, vide infra) setting in at 12 K along with an additional feature (a change of slope) around 5 K for the curves measured in 100 Oe (Fig. 4b); the bifurcation of the low-field ZFC-FC curves in the magnetically ordered state can be seen even for this compound around 12 K. It is important to note that the FC curves in both the cases tend to show an increase with decreasing temperature (though it is weak for the Er case) below $T_N$ which is a signature of cluster spin-glass behaviour [6, 12-14]. The observed values of magnetic ordering temperatures and $\theta_p$ are marginally higher compared to the respective de Gennes scaled values (for full degeneracy), as inferred from the knowledge of the corresponding values (64 K and 86 K) for the Gd analogue [6]. This means that anisotropy of the crystal-field-split 4f orbital probably plays a role on such a breakdown [15].

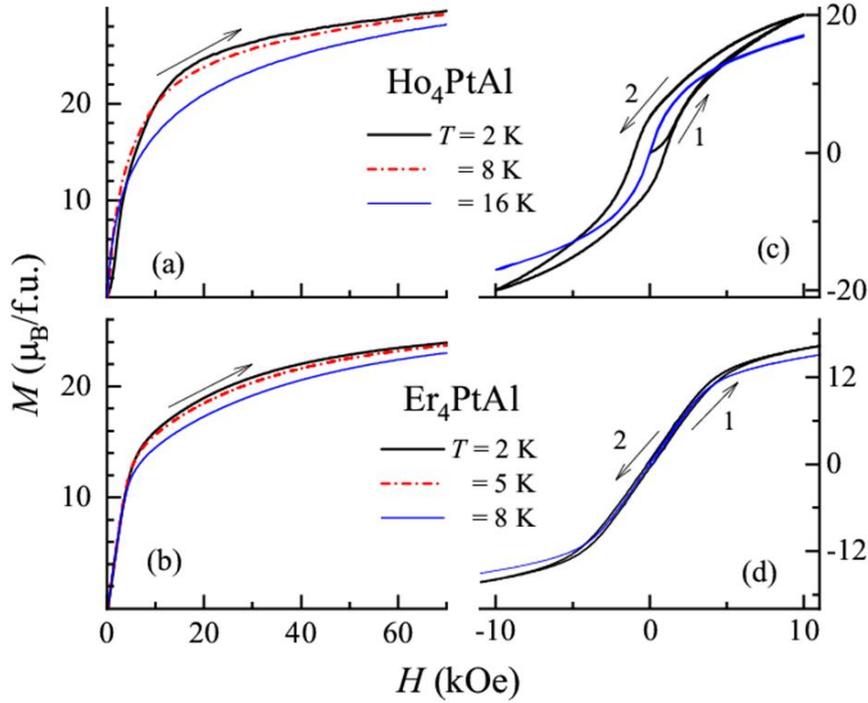

**Figure 5**. Isothermal magnetization (per formula unit) at selected temperatures for (a) $Ho_4PtAl$ and (b) $Er_4PtAl$. In (c) and (d), respective low-field (0 to 10 to -10 to 10 kOe) hysteresis loops are shown for two temperatures.

In Fig. 5a-b, we show the $M(H)$ curves up to 70 kOe at selected temperatures in the magnetically ordered state for both the compounds. The points to be noted are: (i) In these curves, there is a sharp increase of $M$ for initial applications of $H$, indicating possible tendency towards ferromagnetic alignment at a small $H$; there is a distinct step around 5 kOe for the Ho compound at 2 K in support of the fact that the zero-field state cannot be classified as a ferromagnet, but as a canted antiferromagnet undergoing spin reorientation; (ii) even at fields as high as 70 kOe, there is no evidence for saturation, supporting further the canted nature of magnetic structure persisting at high fields; (iii) distinct hysteresis, though weak, is observed for the Ho compound at 2 K, which was found to diminish gradually with increasing temperature well below $T_N$, as shown in Fig. 4c for 16 K, in the low-field hysteresis measurements; this hysteresis suggests the existence of a ferromagnetic component; (iv) hysteresis is absent in the $M(H)$ curves for the Er compound down to 2 K (Fig. 5d). The results overall suggest canted antiferromagnetic nature of the virgin state at the onset of magnetic order.

We have measured ac $\chi$ to get more insight into the nature of the magnetically ordered state, given that the sign of $\theta_p$ is positive supporting the existence of ferromagnetic correlations, whereas the virgin specimens reveal the onset of antiferromagnetic order (as discussed above). The real ($\chi'$) and imaginary ($\chi''$) parts are shown in Fig. 6. In the case of Ho compound, $\chi'$ (in zero field) exhibits a peak at 19 K, followed by a shoulder around 12 K; there are upturns at similar temperatures in the $\chi''$. No feature could be clearly resolved around 5 K, though $\chi''$ exhibits a change of slope. All these features vanish in an applied field of 5 kOe, clearly revealing the existence of a spin-glass component. The $\upsilon$-dependence of the peak, though weak, is discernable; the fact that it is seen even at the onset of long-range magnetic order does not rule out the possibility of antiferromagnetic cluster spin-glass behavior. In the case of Er compound, $\chi'$ (in zero field) exhibits a peak in at ~12 K, followed by a distinct change of slope around 5 K, consistent with the dc $\chi$ data presented above. There is also a shoulder around 10 K, the origin of which is not clear; possibly, there is an additional spin-reorientation at this temperature, which is subtly sensitive to small fields, escaping detection in the dc $\chi$ measurements even in fields as low as 100 Oe. It appears that such subtle additional magnetic features could be observed under favorable circumstances due to complex interaction between the three non-equivalent magnetic sites, as also observed for the Tb analogue [7]. The frequency dependence of the peak is not well resolved, though the left side of the peak shows some dependence. However, the χ" curves reveal a distinct frequency dependent peak at about 5 K without any notable peak at higher temperatures. These findings imply that the 5 K-feature alone could arise from spin-glass freezing. An application of a dc magnetic field of 5 kOe completely suppresses the peaks. Viewing together with the behavior of low-field dc χ curves and dc $M$, we conclude that (virgin state) antiferromagnetism below 5 K behaves like (cluster) spin-glasses for the Er case.

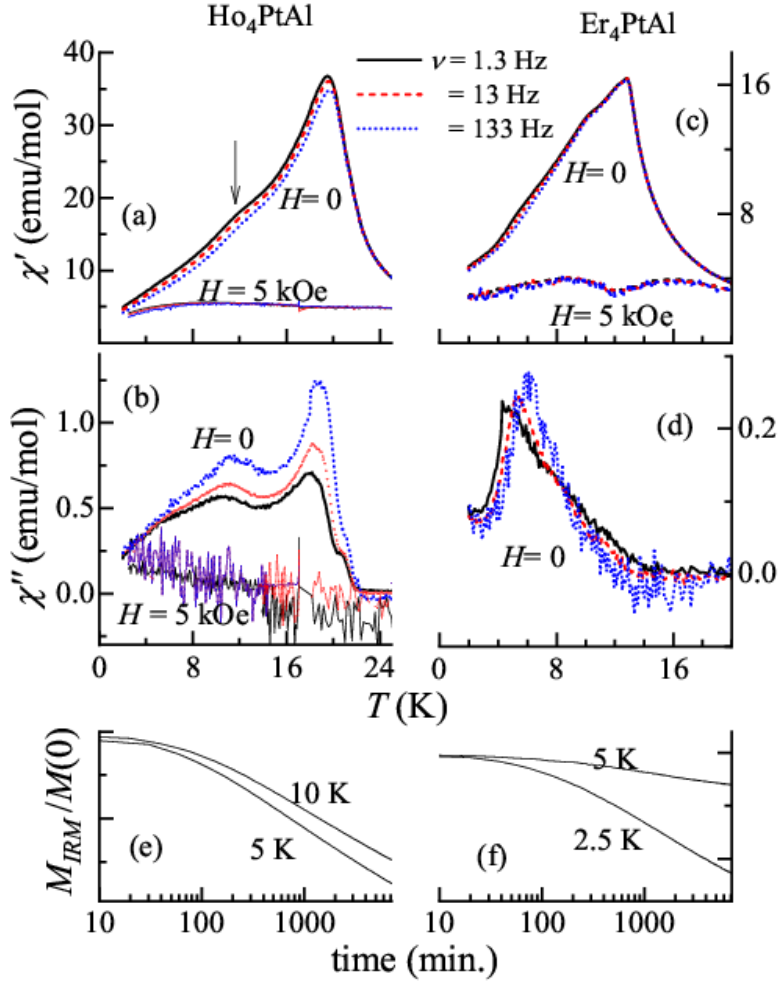

**Figure 6**. Real (χ') and imaginary (χ") parts of ac susceptibility for $Ho_4PtAl$ (left) and $Er_4PtAl$ (right). A vertical arrow in (a) is drawn to show the temperature where the shoulder occurs. Time dependence of isothermal remnant magnetization at selected temperatures are plotted in (e) for the former and in (f) for the latter. In (d), the 5kOe-data is not plotted, as it is very noisy and featureless. The values of $M_{IRM}$, labelled $M(0)$, immediately after the field is switched off, are: For Ho, ~14 and 6.3 emu/g for 5 and 10 K, and, for Er, 2.7 and 1.5 emu/g for 2.5 and 5 K respectively.

In order to render support to the above conclusions for both the materials, we have measured isothermal remnant magnetization ($M_{IRM}$) at selected temperatures below $T_N$; for this purpose, we cooled the samples in 5 kOe to desired temperatures, switched off the field and measured $M_{IRM}$ as a function of time. It is noted (Fig. 6e-f) that the value $M_{IRM}$, measured immediately after the field is switched off, (labelled $M(0)$), decreases with the increase of $T$ and also decreases slowly with time. Though the functional form of the decay with time appears to be complicated (e.g., two logarithmic regions are seen in Fig. 6e-f) possibly due to the presence of multiple sites for $R$, the slow decay is consistent with the glassy behavior, at least in the close vicinity of 5 K for both the cases.

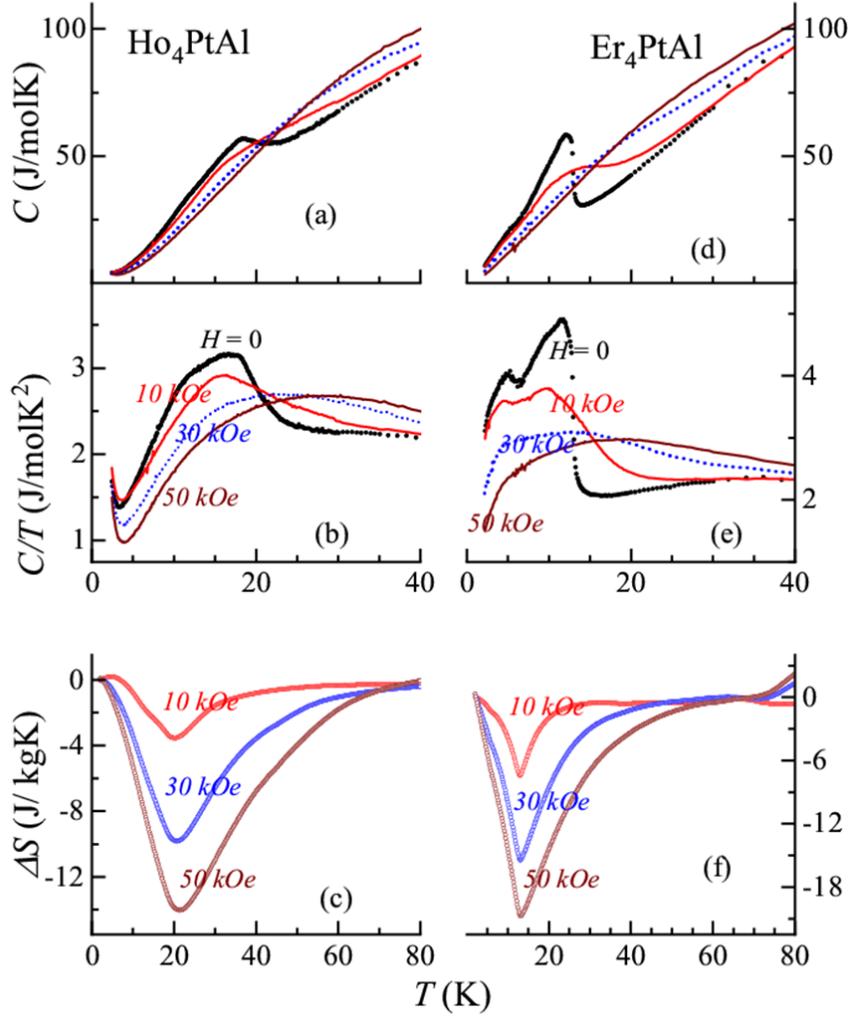

**Figure 7**. The plots of heat-capacity (C), heat-capacity divided by temperature, and isothermal entropy change ($\Delta S$) as a function of temperature for $Ho_4PtAl$ and $Er_4PtAl$.

Fig. 7 shows heat-capacity as a function of temperature for both the compounds in the form of $C$ versus $T$ as well as of $C/T$ versus $T$ below 40 K. It is clear that there is a well-defined feature (upturn followed by a peak) as the magnetic ordering temperature is approached from the paramagnetic state for both the compounds in the absence of a magnetic field. Additionally, for the Ho compound, there is a shoulder around 12 K in the plot of $C(T)$ (Fig. 7a), which is more transparent in the plot of $C(T)/(T)$ (Fig. 7b). There is also a weak upturn in $C(T)$ below about 5 K, and it is possible that it is due to subtle changes in the orientation of magnetic moments. As the magnetic field is applied, for $H$ = 10 kOe, the peak temperature is marginally decreased with the feature due to the onset of magnetic order partly overlapping with the one due to 12 K transition. This establishes that the magnetic structure at the onset of magnetic order is of an AF type, and not of a F-type as proposed earlier [5]. For further higher fields (30 and 50 kOe), the peak is smeared and there is a monotonic decrease of $C$ with $T$ down to 5 K (Fig. 7c). With respect to the Er case, it is clear that, following a sharp rise below 14 K, the peak appears around 12 K (in zero field). The $C/T$ plot (Fig. 7d) shows an additional sharp anomaly around 5 K in support of a magnetic feature around this temperature, as indicated by ac $\chi$. Below 10 K, there is no evidence for $T^{3/2}$ or $T^3$ behavior, expected for ferromagnets/spin-glasses and antiferromagnets respectively, thereby revealing that the magnetism is in fact quite complex. It is not clear whether a shoulder in $C/T$ near 10 K can be correlated to the ac $\chi$ feature at this temperature (see above). The *H*-dependence of the magnetic feature is somewhat similar to that of Ho case and the downward shift of the peak with *H* is consistent with AF ordering at the onset of magnetic transition.

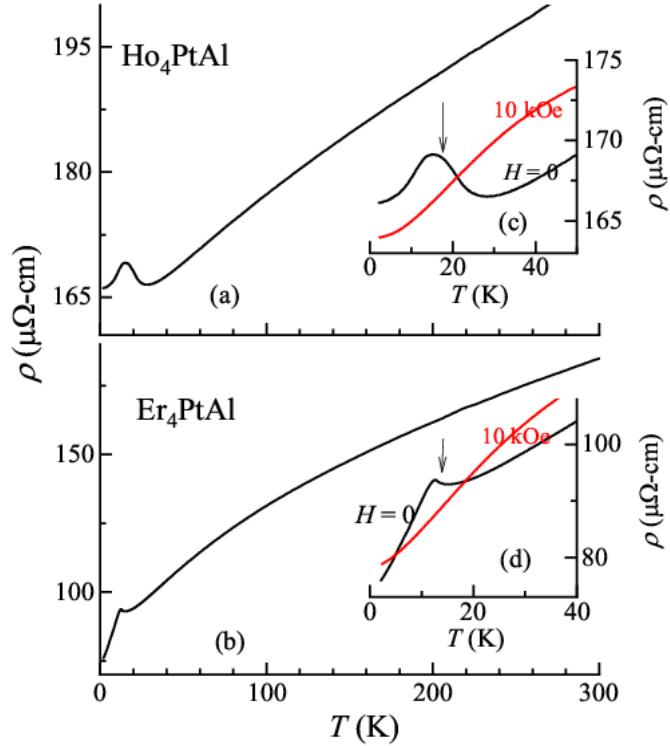

**Figure 8**. Zero-field electrical resistivity in the range 2-300 K are shown for Ho$_4$PtAl and Er$_4$PtAl in (a) and (b) respectively. The data below 50 K for the Ho case and below 40 K for the Er case in zero-field and in 10 kOe are shown in the insets c-f in an expanded form to show the existence of a resistivity minimum. Vertical arrows mark Néel temperature, inferred from other measurements presented in text.

We have derived isothermal entropy change, $\Delta S = S(H) - S(0)$, by integrating the plots of *C/T*, and the results obtained are shown in Fig. 6c and Fig. 6f, for selected final fields. The plots of $\Delta S$ exhibit a peak in the negative quadrant for both the cases, even for an application of a field as small as 10 kOe; this sign is typical of ferromagnetism [16] and therefore this supports the inference made from *M(H)* for the appearance of a ferromagnetic component even for such low fields (though AF component still persists as discussed above). We had also obtained $\Delta S$ from isothermal *M* data employing Maxwells equation for the Er case by measuring isothermal *M* in close intervals of temperature (every 3 K, not shown here); the values and the features are in good agreement with that derived from the *C(T)* data. We would like to emphasize the following outcome: (i) The peak values of $\Delta S$ are relatively large for both the compounds, compared to that of analogous Gd, and Tb members [6-8]. For instance, for a field of 50 kOe, the values for Ho and Er members are about ~14.5 and ~21.5 J/kg K, whereas the corresponding values are lower for Gd (~6 J/kg K). For Tb and Dy cases, the corresponding peak values are ~ 6 and 13 J/kg K respectively. Clearly, the value for the Er sample is the largest within this family. The magnetic refrigeration capacity, defined as the product of the full-width at half maximum and the peak value, is quite large – about 420 J/kg K – and comparable to many best magnetocaloric materials in this temperature range of interest [17]. This finding therefore reinforces [9] that the anisotropy of the (crystal-field-split) 4*f* orbital plays a key role for such an enhancement. (ii) The peak values of these two Pt compounds are far higher than those of the respective member of the isomorphous Rh family as well [9-11]. In fact, a comparison of the peak values of $\Delta S$ between Pt and Rh families suggests that this is in general true for a given *R*, e.g., for Gd$_4$PtAl and Gd$_4$RhAl, the values are ~6 and ~2.3J/kg K respectively [6, 9]. It is known that the electronic correlation within the 4*d* bands is usually stronger than that within 5*d* bands in transition metal systems; on the other hand, the spin-orbit coupling is stronger for the latter. We therefore wonder spin-orbit coupling of the Pt 5*d* orbital plays a role to enhance MCE. (iii) As discussed for other compounds of this family [6, 9-11], the plot of $\Delta S$ exhibits a long tail over a wide $T$-range above $T_N$, possibly arising from a gradual formation of ferromagnetic clusters with a lowering of temperature (behaving like a classical spin-liquid). The readers may note that interesting magnetic precursor effects have been brought out in different context as well in the past in insulting materials [18]. (iv) The isothermal magnetization curves are non-hysteretic for the Er case, and, in view of this, this compound can be added to the list of materials for magnetic refrigeration below 40 K, considering that it is often emphasized in the literature to find materials with such a reversible behavior [19, 20].

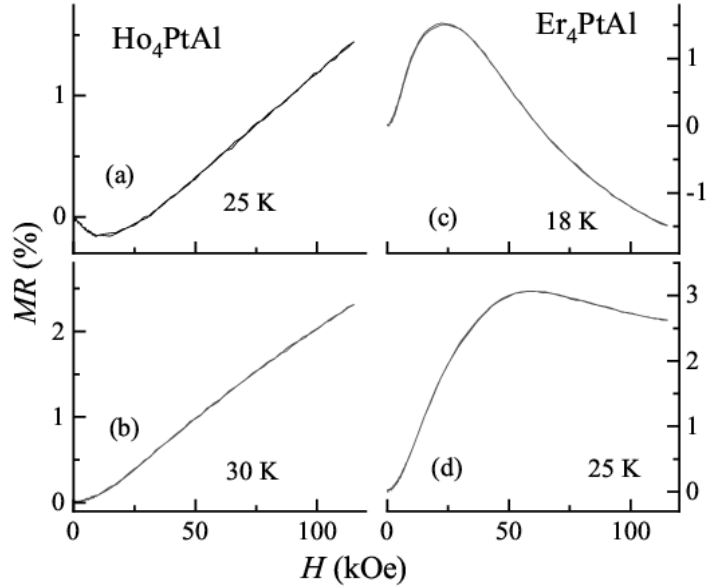

**Figure 9**. Isothermal magnetoresistance [$\rho(H)-\rho(0)$]/$\rho(0)$] as a function of magnetic field at selected temperatures above $T_N$ for Ho$_4$PtAl and Er$_4$PtAl.

The $\rho(T)$, as expected for metals, exhibits positive temperature coefficient as the temperature is lowered below 300 K (Fig. 8a-b), but exhibits a minimum in the paramagnetic state just above the ordering temperature in both the cases in zero field (at 26 K and 14 K, respectively) (Fig. 8c-d, insets). Such a minimum, though not expected for rare-earths with localized 4$f$ electrons like Ho and Er, was encountered in some heavy rare-earth systems as a precursor to long range magnetic order [10, 21, 22]. Recent new theoretical ideas attribute it to classical spin-liquid behavior due to geometrically frustrated magnetism [23]. Such a minimum vanishes gradually with an application of external magnetic field, as shown in Figs. 8c-d, for an application of 10 kOe. As the material enters magnetically ordered state, the drop due to the loss of spin-disorder contribution (see, zero field curves) can be seen. Following interesting observations are made on the magnetoresistance (MR= [$\rho(H)-\rho(0)$]/$\rho(0)$] data, even in the paramagnetic state (Fig. 9). There is a competition between positive contribution due to the Lorentz motion of the conduction electrons (that is, classical metallic part, varying quadratically with $H$) and negative contribution due to the suppression of spin fluctuation contribution by $H$ as a function of $H$ and $T$. Thus, at 18 K (close to $T_N$), for the Er case, there is initially an increase in the positive quadrant due to metallic contribution, followed by a peak around 25 kOe and then a fall with the curve entering the negative quadrant around 55 kOe due to spins. At higher temperatures, say at 25 K, MR curve remains in the positive quadrant due to weakening of spin contribution with respect to metallic part. On the other hand, for the Ho compound, just above its $T_N$, say, at 25 K, the spin contribution dominates as revealed by the negative sign till about 25 kOe and it is overcompensated at higher fields by the metallic part at higher fields. At higher temperatures, the MR($H$) curve stays in the positive quadrant only. With respect to the behavior below $T_N$ (Fig. 10), the sign of MR remains negative which is either due to antiferromagnetic gap formation and/or spin-glass component. The fact that the antiferromagnetic-gap formation occurs is evidenced, at least for the Ho case, by the observation that there is initially a (weak) upturn in $\rho$ (in the zero-field curve) as soon as the magnetically ordered state is entered; this upturn is suppressed by a field of 10 kOe (see Fig. 8). A notable finding in the isothermal MR curves is that the virgin curve lies prominently outside the envelope loop below 12 K in the low-field range (< 4 kOe) with a significant hysteresis for the Ho case, supporting that there could be a disorder-broadened first-order magnetic transition at such low fields; however, in the Er case, the magnitude of MR is so small that hysteresis MR($H$) can be considered negligible.

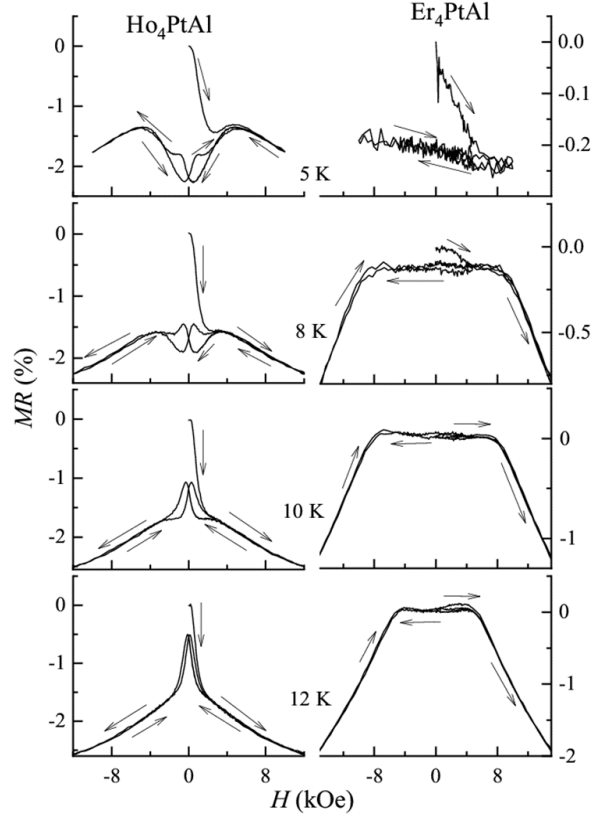

**Figure 10**. Magnetoresistance hysteresis loops at 5, 8, 10 and 12 K for Ho$_4$PtAl and Er$_4$PtAl.

**4. Conclusions**

The present results establish that Ho$_4$PtAl and Er$_4$PtAl, are interesting magnetic materials with re-entrant spin-glass behaviour with the onset of magnetic order of an antiferromagnetic type, but undergoing subtle changes in this magnetic state by an application of small external field. A notable finding is that these compounds show the largest value of isothermal entropy change (a measure of magnetocaloric effect) at the onset of magnetic order within this family, in particular with respect to the isomorphous Gd compound. Since the values surpass that of a S-state ion, this finding suggests that topology of the 4$f$ orbital can enhance magnetocaloric effect. Another intriguing outcome, based on the comparison of MCE behavior of Pt based family with that of Rh family, is that Pt 5$d$ spin-orbit coupling also may play a role in this regard. We hope these inferences provide clues for the advancement of theories in the field of MCE to enable engineering of materials for magnetic refrigeration at room temperature. This entropy behaviour of the Er compound meets the much needed [20] characteristic of 'reversibility' for magnetocaloric applications in the low temperature range. Finally, neutron diffraction studies would be rewarding to understand the magnetic structure changes with temperature and magnetic field in these materials.

**Acknowledgments:** E.V.S. thanks Atomic Energy Department, Government of India, for awarding Raja Ramanna Fellowship. Financial support from DAE, Govt. of India (Project Identification No. RTI4003, DAE OM No. 1303/2/2019/R&DII/DAE/2079 dated 11-02-2020) is thankfully acknowledged. K. M. thanks financial support from BRNS, DAE under the DAE-SRC-OI program. S. M. thanks support from Vision Group on Science and Technology-GRD No 852.